\documentclass[%
 reprint,
 amsmath,amssymb,
 aps,
 showkeys,
]{revtex4-1}

\usepackage{graphicx}
\usepackage{dcolumn}
\usepackage{bm}
\usepackage{hyperref}
\usepackage{tabularx}

\usepackage{xcolor}
\usepackage{colortbl}
\usepackage{multirow}
\newcolumntype{b}{>{\centering}X}
\newcolumntype{s}{>{\hsize=.6\hsize\centering}X}
\newcolumntype{B}{>{\hsize=2\hsize\centering}X}

\usepackage{caption}
\usepackage{subcaption}

\begin{document}

\title{Locating the source of large-scale diffusion\\of foodborne contamination} 

\author{Abigail L. Horn}
 \email{abbylhorn@alum.mit.edu}
 \affiliation{%
 Institute for Data, Systems, and Society, Massachusetts Institute of Technology\\
 77 Massachusetts Ave, Cambridge, MA 02139}
 \altaffiliation[Present address: ]{Federal Institute for Risk Assessment (BfR),
 Max-Dohrn-Stra\ss{}e 8-10, 10589 Berlin, Germany}

\author{Hanno Friedrich}%
\affiliation{%
 K\"uhne Logistics University\\
 Gro\ss{}er Grasbrook 17, Hamburg 20457\\
}%

\date{\today}

\begin{abstract}
We study the problem of identifying the source of emerging large-scale outbreaks of foodborne disease. To solve the source identification problem we formulate a probabilistic model of the contamination diffusion process as a random walk on a network and derive the maximum likelihood estimator for the source location. By modeling the transmission process as a random walk, we are able to develop a novel, computationally tractable solution that accounts for all possible paths of travel through the network. This is in contrast to existing approaches to network source identification, which assume that the contamination travels along either the shortest or highest probability paths. We demonstrate the benefits of the multiple-paths approach through application to different network topologies, including stylized models of food supply network structure and real data from the 2011 EHEC outbreak in Germany. We show significant improvements in accuracy and reliability compared with the relevant state-of-the-art approach to source identification. Beyond foodborne disease, these methods should find application in identifying the source of spread in network-based diffusion processes more generally, including in networks not well approximated by tree-like structure.

\end{abstract}

\keywords{Complex systems, Source identification, Epidemic, Spreading, Diffusion, Network Diffusion, Food supply networks, Foodborne Disease}

\maketitle

\section{\label{sec:intro}Introduction}

The complexity and globalization of food production have made foodborne disease a widespread public health problem in both developed and developing countries. Most outbreaks of foodborne disease involve a source of contamination at the point of preparation or sale and affect a small group of people in a localized area. However a small but worrisome minority of outbreaks are generated by a contamination originating at the site of production or processing, generating a widespread diffusion of contamination through the supply chain and affecting a potentially much greater number of people across geographically distributed locations. When large-scale outbreaks do occur the impact on the public's health may be massive. In the summer of 2011 an outbreak caused by Shiga toxin-producing \textit{Escherichia coli} (EHEC) O104:H4, spread by sprouts grown in Germany, caused 54 deaths and 4,321 illnesses in 16 countries over a nine-week period \cite{1,2}. As the food system continues to become interconnected, driven by large-scale production practices and distribution over ever-larger distances, both the prevalence and the severity of consequences of large-scale outbreaks are increasing. From 2005 - 2014, nearly 200 multi-state outbreaks were identified and investigated in the US as compared with 85 over the years 1995 - 2004; these multi-state outbreaks accounted for 3\% of total outbreaks, but were responsible for 34\% of hospitalizations and 56\% of deaths \cite{3}.

In the event of a large-scale outbreak, rapidly identifying the contamination source is essential to minimizing impact on public health and industry. There are three standard components to the regulatory response and investigation process, each contributing to the challenge of identifying the source: (i) detecting that an outbreak is occurring, (ii) identifying the food vector causing the outbreak, and (iii) identifying the location source of the outbreak at a farm or processing center. Novel strategies facilitated by new analytical tools and the increasing availability of data sources are being developed to improve to the ability to (i) detect outbreaks, e.g. by crowdsourcing self-reported foodborne illness concerns from popular social networking sites  \cite{4,5,6} and (ii) implicate the food type or even specific product carrying the disease, e.g. by analyzing retail-scanner data from grocery stores \cite{7,8}. This paper addresses part (iii) of the outbreak investigation, identifying the location of origin.

Tracing the location of the source of an outbreak is a challenging problem due to the complexity, dynamics, and massive scale of the food supply network and the absence of integrated labeling and distribution records. However, current investigation methods represent a missed opportunity to utilize valuable information to solve the source localization problem. The regulatory approach generally involves \textit{triangulation}, or tracing back the unique distribution paths of products from several locations to determine if there is a point of convergence in the supply network, such as a common date and location of harvest or place of manufacture \cite{9,10,11}. Because of resource limitations, investigators are only able to make use of a small subset of the reported cases of illness -- data that serves as evidence in the source location problem. With only a few pieces of evidence, the time consuming traceback will often be unsuccessful in narrowing down the problem significantly. As a result, investigations are completed in many cases after the outbreak has ended and the contamination has made its way through the supply network, meaning that no cases of illness are averted.

Furthermore, the majority of outbreaks remain unsolved, meaning that the food and/or location source of the outbreak is never identified \cite{11,12}. In the 13,352 foodborne disease outbreaks (causing 271,974 illnesses) documented by the US Center for Disease Control and Prevention (CDC) during 1998-2008, only 4,887 (37\%) were traced back to a single food vehicle and pathogenic source, with less than 15\% of these to a specific contamination point \cite{13}.

\subsection{\label{sec:netapproach}A network approach to source identification}

Food distribution is a complex system that can be seen as a network of trade flows connecting supply network actors. Identifying the source of an outbreak of contamination distributed across a network can best be solved by considering this network structure and the dimensions of information it contains. Together with reports of illness, this network information can be used to better solve the problem of identifying the source of large-scale outbreaks.

The literature on the network source identification problem has grown widely in recent years covering problems in many different contexts, from contagious disease infecting a human population, to computer viruses spreading through the Internet, to rumors or trends diffusing through a social network. Much of this work has focused on studying this problem in analytically tractable frameworks, designing approaches to work on trees and extending to general network structures in an \textit{ad hoc} manner \cite{14,20,21,22,23,24,25,26,27}. These simplified frameworks lack many features of real-world networks and problem contexts that can dramatically impact transmission dynamics, and therefore, backwards inference of the transmission process. Moreover, the features that distinguish foodborne disease in the context of source identification have not previously been studied or identified. We review these features in Section~\ref{sec:ProblemFraming} and conclude that much of the existing work cannot be implemented in the foodborne disease problem because it makes assumptions about the transmission process that are unrealistic in the context of food supply networks -- that is, identifying the source of an epidemic \textit{contagion} whereas foodborne contamination spreads through a transport-mediated \textit{diffusion} process \cite{18,19,22,23,24,25}, or because it requires data that is not available -- complete observations of the contamination status of all nodes in the network \cite{14,15,16} or timed network data \cite{18,19,20,21,22,23}. 

\subsection{\label{sec:contributions}Problem statement and contributions}

To solve the source detection problem in the context of foodborne disease, we assume as given a network model of the supply of a specific food commodity and a probabilistic model of the transmission process of contamination spreading through this network. At some point in time a single contamination source begins to send out contaminated products, which travel through the network according to the transmission model, resulting in in observations of illness at a set of network nodes. Our objective is to minimize the error between our estimate of the location of the source and the true location of the source in the network, given the locations of the observations of illness.

We formulate the transmission process of contaminated food items traveling through the supply network as a discrete Markov chain, i.e. a random walk on the network where transmission probabilities correspond to the edge weights. This is a natural transmission model for non-contagious diffusion on a weighted, directed network. To estimate the true source location we adopt a maximum likelihood (ML) approach, by definition minimizing the estimation error. The ML estimator chooses the highest probability source node according to the likelihood of observing the reports of illness.

By formulating the transmission process as a weighted random walk, we are able to develop a computationally tractable representation of the ML estimator that accounts for all possible paths of all possible lengths traveled by each contaminated food item. This is in contrast to existing approaches to source detection in weighted, time-homogeneous networks, which develop heuristic methods that consider only the shortest path \cite{24,25}, or the dominant (i.e. highest-probability and shortest) path \cite{26,27} between each source and each observation. While such approximations may be justified in other network contexts, considering only one possible path of travel between each source and each observation can dramatically limit the ability to model the transmission dynamics of foodborne disease outbreaks, and therefore reduce the accuracy of source detection.

Practically, we demonstrate that a source estimator that accounts for all possible transmission paths can locate the outbreak source with greater timeliness, accuracy, and reliability than other methods, and more robustly across extreme cases of network structures. This is shown through the application to different network topologies, including stylized models of food supply network structure and simulated outbreaks, and real network and illness data from the 2011 EHEC contaminated sprout outbreak in Germany. Compared with the relevant state-of-the-art approach \cite{26,27}, the improvement in accuracy with our source estimator is always observable and can be substantial, ranging from a 10\% to 80\% improvement depending on the network topology evaluated. Furthermore, application to the EHEC outbreak demonstrates that our approach is not only more accurate but also more reliable, consistently identifying the source location region over the time course of the outbreak.

Methodologically, though our paper is motivated by the case of foodborne disease, the ML source estimator developed here can be extended to identify the source of network-based diffusion spreading processes more generally. Because these methods are not limited to tree-like network approximations, they should find application in a greater range of problem contexts and demonstrate similar benefits in accuracy and reliability of source localization as those shown for foodborne disease.

\subsection{\label{sec:outline}Outline}

The remainder of this paper is organized as follows. In Section~\ref{sec:ProblemFraming}, we review the existing literature on the network-based source detection, first outlining key features that distinguish this problem from source detection in other network contexts then using these features to categorize the existing approaches according to their relevance to the foodborne disease source detection problem. In Section~\ref{sec:Model} we introduce the probabilistic foodborne disease transmission model and derive the source estimator. In Section~\ref{sec:Evaluation} we demonstrate the effectiveness of our source estimation framework through application to both stylized networks and real data from 2011 EHEC outbreak. In Section~\ref{sec:Conclusions} we conclude and discuss future work.

\section{\label{sec:ProblemFraming}Problem Framing and\\ Related Work}

To formulate the problem of source detection on a network, assumptions must be made regarding (i) the network and observation data available for source identification, and (ii) the transmission process that led to the observations. Based on basic practical knowledge of food supply networks and the foodborne disease contamination process, we introduce the source identification problem in the context of foodborne disease outbreaks and outline six features that distinguish this problem from source detection in other network contexts due either to practical data limitations or differences in transmission process mechanics. We then use the six features to categorize the existing literature on the network-based source detection problem according to relevance to the foodborne disease context.

\subsection{Background and Definitions}

Food supply systems can be represented by a directed network structure consisting of multiple stages of production, distribution, storage, and consumption. Flows through the network are generally structured such that product is distributed in a forward direction along a \textit{path}, or a collection of directed \textit{edges} connecting supply \textit{nodes} from origination to point of sale. A large-scale outbreak occurs when contaminated food departs from some source in an early stage of the network that is able to reach downstream nodes in geographically distributed locations. The contamination will eventually make its way to consumers, who develop illness some time after consuming the contaminated food. Case reports of illness are associated with the supply network node at which the offending product was purchased and \textit{exits} the supply network, e.g. a retailer or restaurant; these nodes can be considered \textit{infected}.

\begin{figure}[h]
\includegraphics[width=.5\textwidth]{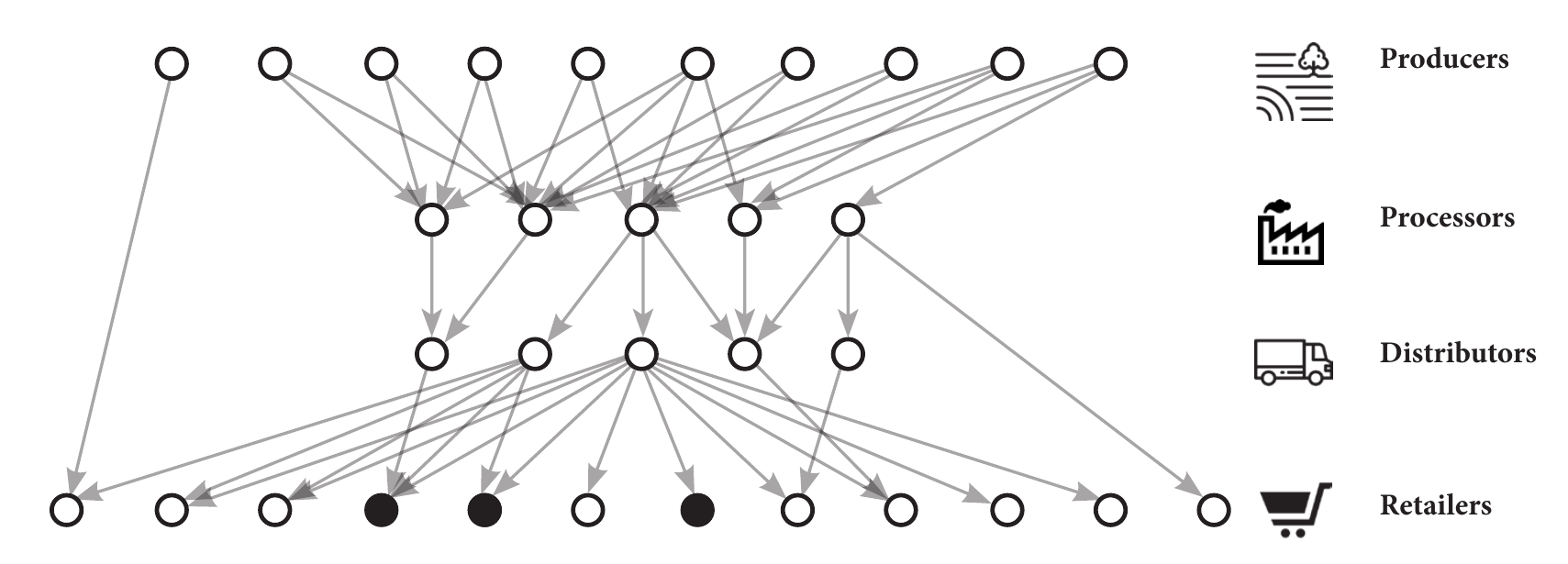}
\caption{\label{fig:01}Illustration of a food distribution network with three reported cases of illness (at the shaded nodes) linked to retailer nodes.}
\end{figure}

The network in Fig.~\ref{fig:01} represents a supply network in which contamination at a food producer has spread through the supply network, leading to reports of illness at 3 different retailers. With this structure mapped, it is straightforward to utilize all case data (i.e. evidence) available during an event to identify the set of \textit{feasible} sources of contamination, that is, the set of nodes that connect to all known contaminated nodes. Network structural information thus provides a first cut into the source identification problem by enabling the identification of feasible sources. To differentiate between the feasible sources, further dimensions of information available within the network can be leveraged. Each edge contains information about the volume of goods traded between supply network actors. Volume-weighted information is a source of heterogeneity that can be thought of as the relative propagation potential of a given edge, providing insight into the paths along which contaminated product is likely to have traveled.

\subsection{Distinguishing features of\\ foodborne disease transmission}

\subsubsection{A transport, not epidemiological, transmission process}

Many network-based source detection methods are designed to identify the source of an infectious contagion. These methods often assume some variant of the epidemiological model of contagion transmission, including the widely used susceptible-infected (SI) or susceptible-infected-recovered (SIR) models. However the transmission of contamination through the food supply to people is different from the disease contagion process from people to people. As contaminated (solid, perishable) food moves through the supply network, the pathogenic quantity will generally remain conserved, meaning it will neither spread to other food items nor decay significantly in infectivity \cite{28,29}. The former is due to a number of factors including the lack of contact between packaged items, the lack of interaction or mixing between unpackaged items, and the biological insusceptibility of contamination to spread. The latter is due to the shelf life of perishable items being shorter than the pathogenic decay rate. The contamination process in the context of foodborne disease therefore largely involves the contaminated food \textit{diffusing} rather than \textit{contagiously spreading} through the supply network. Transmission models based on the epidemiological process of contagion and recovery will add complexity without representing additional dynamics of the diffusion process.

Additionally, the observation data available for source identification occurs on the human level and only via infection status, (I) in the SI/R model. Observations of contamination occur when people report illness. Each illness is linked to a supply network node at which the contaminated food was purchased. Data regarding the contamination status of individual food items is not normally available during an investigation. Furthermore, it is not possible to establish from the illness reports whether a supply node has ever received contaminated food and is thus susceptible (S), as it may have led to illnesses that went unreported. Methods that rely on observations of susceptible status or that assume nodes not reporting infection are contamination-free (also called ``negative information'') are thus non-applicable in this setting.

\subsubsection{Observations are sparse}

Though the contamination will travel through multiple network nodes on its journey through the supply network, it is only observed when illness is reported in connection with the exiting or \textit{absorbing} node at which contaminated food was purchased. The contamination status of \textit{transient} nodes involved in the production, processing, or storage of food, though closer to the source in number of network edges, will remain hidden to investigators unless further investigations are performed (normally during later stages of an investigation). A trivial implication of the sparsity of observations is that it is unrealistic to assume, as some source detection methods do, that the contamination status of all nodes in the network is known.

\subsubsection{Observations will always be spaced far from the source}

The placement of observations only at absorbing nodes also means that there will be a large network distance between the source and each observation, increasing the number of possible paths that could have been traveled and in turn the uncertainty in the \textit{structure} of the diffusion trajectory. At the same time, the differing volume-weights along the edges of the supply network provide valuable information for inference. Given the large uncertainty in the diffusion structure, approaches to source detection that consider network structure alone will be inferior to those that consider this weighted information.

\subsubsection{Similar path lengths}

Due to the staged structure of the food supply network, paths through the network from source to observation will be close to the same length in terms of number of network edges. Many existing source detection methods simplify the inference process by assuming that the contamination traveled across the shortest path from the source to each observation, or otherwise by leveraging shortest path properties of graphs. These approximations will apply poorly in the food supply network context where most paths will be indistinguishable in length.

\subsubsection{Multiple candidate paths}

Between any possible source and observation in a food supply network, there exist multiple paths of travel of similar weight or likelihood. This is due to the lack of monopolies in food production, trade, and retailing markets: any given food type will be distributed through multiple larger retailers or wholesalers, each dealing with similarly large volumes of product. Certain source detection methods make the simplifying assumption that the contamination travels across the single highest-probability path between a source and observation. These methods will be inaccurate in the food supply network setting where transmission dynamics are not necessarily dominated by a small percentage of connections.

\subsubsection{Data on times through the network are lacking}

In theory, there should be a signal for source detection from the timed reports of illnesses combined with a model of the time it takes to transmit the contamination. Each collection of edges in a network path encodes information about the time delay that a contaminated product could have taken to travel these steps. These delays will be distributed differently according to parameters like the distance and speed of travel and supply network logistics encountered. However, there is significant temporal uncertainty in the contamination transmission process. The time the contamination may spend in storage, both at various nodes along the supply network (e.g. warehouses) and with the consumer after purchase, as well as during the incubation period, can be significant and vary widely -- and potentially much more so than the time spent in travel. Furthermore, while the times of infection are available to some degree of accuracy (recorded according to patient recalled time of illness onset), data on storage times through the network are unavailable with the exception of a few case-specific customer or retailer survey studies \cite{29,30}. Therefore, while time can be an important aspect in some foodborne disease source detection applications, time-based methods are not currently implementable in the foodborne disease context given available data.

\subsection{Categorization of literature}

Many approaches to the network source detection problem have been developed in recent years, though none of these methods have specifically considered the context of outbreaks of foodborne disease. We now briefly review the major themes in the existing work, using the features described above to guide the discussion in terms of relevance to the problem on food supply networks. These features are summarized in Table ~\ref{tab:table1}.

The earliest approaches to source detection are based on complete observations, relying on knowing the contamination status (SI/R) of each node in the network a fixed point in time \cite{14,15,16}. These methods do not incorporate information about differing weights along edges but are based solely on graph structure by employing notions of network centrality, the intuition being that the node most ``central'' to the observed contamination process is the source. The seminal work by Shah and Zaman \cite{14} introduces the measure of \textit{rumor centrality}, which considers the number of linear extensions between each source and the infected nodes. The method and analytical results concerning detection probability are derived for trees or tree-like graphs; to apply to general networks, a Breadth-First-Search (BFS) heuristic that assumes the contamination traveled across the shortest paths to the observations must be used. Other methods based on \textit{betweenness centrality} \cite{15} and \textit{eigenvector centrality} \cite{16,17} apply to general networks without employing a shortest path heuristic, although the calculation of betweenness is based on shortest path properties. These methods were important for establishing foundational results on the network source detection problem, but are impractical due to the complete observation assumption.

Many methods have since been developed for the more realistic setting that only a subset of the contaminated nodes are observable, i.e. partial observations. These can be categorized into temporal methods -- approaches designed to make use of the information from the timed reports of illness and times through the network, and non-temporal methods -- approaches that rely only on the node location where contamination has been reported. The temporal category includes methods assuming discrete-time epidemic (SI/R) contagion models based on \textit{dynamic message-passing} \cite{18} and \textit{Bayesian belief propagation} \cite{19} equations, or continuous-time \textit{Gaussian} transmission models \cite{20,21}. While a continuous-time transmission model is a better approximation for realistic settings, this approach in \cite{20,21} is limited by being designed for trees and extended to general graphs via a BFS (shortest-path) heuristic. Other temporal methods have been proposed that observe the contamination status of a subset of sensor nodes at user-controlled intervals \cite{22,23}. These methods are impractical for the foodborne disease context given the lack of temporal data available for solving the problem. Furthermore, the methods based on contagion models \cite{18,19,22,23,24,25} are inapplicable in the case of the transport-mediated diffusion process of foodborne disease spread.

Fewer approaches to source detection exist within the category of non-temporal approaches based on partial observations. A line of work based on the notion of \textit{Jordan centrality} has led to multiple variants of a technique that chooses the source node with the shortest maximum path length over all observations, that is, the Jordan center \cite{24,25}. While this method has been extended to incorporate weights along the edges\footnote{In the contagious disease context, weights can be interpreted as heterogeneous infection probabilities.} \cite{25}, it relies on path lengths to discriminate between sources. Furthermore, the technique is designed for tree-like networks; for application to general topologies a separate procedure based on closeness centrality (i.e., counting the sum of the shortest path to each observation) is proposed.

Another line of work in the category of non-temporal approaches involves a measure of \textit{Effective Distance} on a network, defined such that the shortest, highest probability path from a source to an observation has the shortest Effective Distance through the network \cite{26,27}. To identify the source of an outbreak, the single shortest Effective Distance path to each observation is identified. The source is then chosen as the node that minimizes the average and variance of the shortest Effective Distance to each observation.

The Effective Distance method is the existing source detection approach most relevant to the foodborne disease setting and has been evaluated in application to the 2011 outbreak of EHEC in sprouts \cite{27}, also considered as an application case in Section~\ref{sec:Evaluation}. Nonetheless, it is a heuristic approach that considers only a single path to each observation. It was designed for application to global mobility networks which are characterized by great heterogeneity in path lengths and probabilities, and for which contamination processes will be dominated by a small percentage of the shortest, highest probability transport connections. It performs well in those settings, i.e. outbreaks of infectious disease (e.g. SARS, H1N1) mediated through global air travel networks \cite{26}. In food supply networks where path probabilities are less differentiated the Effective Distance method will consider only the single, highest probability path, ignoring the contribution of plausible paths of marginally lower probability. The unsuitability of this heuristic in the food supply network context is observable when applied to the 2011 EHEC outbreak, where source identification results are unstable and less accurate than the infectious disease case examples \cite{26,27}.

\subsection{Summary}

Many existing approaches to the source detection problem cannot be implemented in the foodborne disease context because they are designed for a different purpose -- identifying the source of an epidemic contagion \cite{18,19,22,23,24,25} whereas foodborne disease is spread according to a diffusion process, or because they require data that is not realistically available -- complete observations of the contamination status of all nodes in the network \cite{14,15,16} or timed network data \cite{18,19,20,21,22,23}. Those that are implementable are limited by unrealistic assumptions regarding the transmission process. These methods apply tree-like approximations to deal with general graphs, assuming contamination always travels from source to observations along the shortest \cite{24,25} or highest probability \cite{26,27} paths. While this type of approximation is justified in certain network contexts, food supply networks are not well approximated by tree structure. Moreover, these methods are by definition approximations that do not explore the full set of trajectories between each source and observation. The random walk transmission model developed in this work aims to address this limitation, presenting a computationally tractable approach to calculate the total probability of traveling between a source and each observation along all possible paths of all possible lengths. The resulting approach is not only relevant for solving the source identification problem in food supply networks but also represents a methodological improvement for source identification in diffusion processes more generally.

\begin{table*}
\caption{\label{tab:table1}Categorization of existing work on the source detection problem according to relevance to the foodborne disease context.}
\resizebox{\linewidth}{!}{\begin{tabularx}{1.25\textwidth}{|b|s|s|s|s|s|s|}
\cline{1-7}
 \cellcolor{gray!50}& \multicolumn{6}{c|}{Limitations of source identification methodologies in foodborne disease context}\tabularnewline\cline{2-7}
 \cellcolor{gray!50}& (1) &(2) &(3) &(4) &(5) &(6)\tabularnewline\cline{2-7}
\multirow{-3}{3cm}{\cellcolor{gray!50}Source identification methodology of existing work} & Assumes SI(R) model/status &Assumes
complete observations &Ignores weights &Only shortest paths &Only dominant paths &Assumes observation times\tabularnewline\cline{1-7}
\cellcolor{gray!50}Rumor centrality\\ \cite{14} & & X & X & X & & \tabularnewline\cline{1-7}
\cellcolor{gray!50}Betweenness centrality \cite{15} & & X & X & X & & \tabularnewline\cline{1-7}
\cellcolor{gray!50}Eigenvector centrality \cite{16,17} & & X & X &  & & \tabularnewline\cline{1-7}
\cellcolor{gray!50}Message passing \cite{18} & X &  &  &  & & X \tabularnewline\cline{1-7}
\cellcolor{gray!50}Belief propagation \cite{19} & X &  &  &  & & X \tabularnewline\cline{1-7}
\cellcolor{gray!50}Gaussian\\ \cite{20,21} & &  &  & X & & X \tabularnewline\cline{1-7}
\cellcolor{gray!50}Four-metric\\ \cite{22} & X &  &  & X & & X \tabularnewline\cline{1-7}
\cellcolor{gray!50}Monte Carlo\\ \cite{23} & X &  &  & X & & X \tabularnewline\cline{1-7}
\cellcolor{gray!50}Jordan centrality\\ \cite{24,25} & X &  &  & X & &  \tabularnewline\cline{1-7}
\cellcolor{gray!50}Effective Distance \cite{26,27} &  &  &  &  & X &  \tabularnewline\cline{1-7}
\end{tabularx}}
\end{table*}


\section{\label{sec:Model}Source Detection Model}

\subsection{Problem Statement}

Our goal is to identify the source of a foodborne disease outbreak based on the reports of illness and information about the underlying network structure. We assume as given a network model and a probabilistic model of the transmission process of contamination spreading through this network. At some point in time a single contamination source $s^*$ begins to send out contaminated products, which spread through the network according to the transmission model, resulting in a list of observations of illness $\Theta$ associated with a set of network nodes $O$. Our objective is to minimize the error of our estimate of the source location in the network and the location of the true source, given the information from the observations of illness. To estimate the true source location we adopt a maximum likelihood (ML) approach that chooses the highest probability source node according to the likelihood of observing the reports of illness, by definition minimizing the estimation error.

In the following we describe our model of the food supply network and the foodborne disease transmission process. We then define the ML source estimator for food distribution networks.

\subsection{Food supply network model}

We model the food supply network as a directed graph $G= \{V,\,E\}$, where $V$ is the set of nodes representing supply network actors. $G$ consists of two types of nodes: the set of absorbing nodes $V_R$ and the set of transient nodes $V_Q$, such that $V=\{V_Q,\,V_R\}$. Absorbing nodes represent the point at which product is purchased for consumption and departs the supply network, never to reenter (e.g. retailers or restaurants). All other nodes are transient, representing the points at which food is generated or produced, processed, and stored. $E$ is the set of edges of the form $(i, j)\in V_Q \times V_Q \cup V_Q \times V_R$, representing trade relationships. Each edge $(i, j)$ is weighted by the volume of food shipped over a certain time period from $i$ to $j$, $w_{ij}$.

\subsection{\label{sec:transmissionModel}Foodborne disease transmission model}

The process leading to foodborne disease illness presentation consists the initial inoculation of contaminated product somewhere in production and subsequent dispersal through the food supply network, followed by the transmission of contamination from product to person, ending in illness.

At the core of our model of this process are the following assumptions:

\begin{itemize}
\item[i.] The contaminated quantity is fixed, and is composed of individual contaminated units that neither spread nor recover from contamination as they travel through the supply network.
\item[ii.] Each contaminated unit travels independently through the supply network.
\item[iii.] Each transition of a unit from one node to the next entails an independent transmission direction.
\end{itemize}

Taken together, these assumptions describe a unit-centric diffusion process that can be visualized as a large number of ``pinballs'' traveling through a pinball machine, each ball's trajectory determined according to a stochastic process described below.

Given these assumptions, we introduce the diffusion model. First, some initial quantity of product produced at a \textit{single} unknown source $s^*$ is contaminated. We model $s^*$ as a random variable with a predefined prior probability distribution, $P(s^* = s)$ over $s \in V_Q$ . The contamination diffusion process is initiated when batches or units of
contaminated product depart from $s^*$.

Due to the second assumption, a discrete-time Markov process determines the movement of a contaminated unit,
i.e. a weighted random walk through the supply network. The sequence of states $X_n$ obtained in successive transitions are determined by the state-to-state Markov transition probabilities,
\[
P(X_{n+1}=j|X_n =i)=p_{ij}.
\]

The $p_{ij}$ taken together compose $P$, the row-stochastic Markov transition matrix for the transmission process of a contaminated unit occurring on network $G$. The probability of self-transition is defined as $p_{ii} = 1$ for absorbing nodes $i \in V_R$ . Because of the supply network structure, it is convenient to consider $P$ in an aggregated, ordered form such that the absorbing nodes come last. We can then unite the transient nodes and the absorbing nodes so that the form of the transition matrix becomes
\[
P=\left[\begin{array}{cc}
P_Q & P_R\\
0 & I_R
\end{array}\right],
\]
where $P_Q$ is the $|V_Q| \times |V_Q| $ sub-matrix concerning transitions between transient nodes, $P_R$ is the $|V_Q| \times |V_R|$   sub- matrix concerning transitions from transient nodes into absorbing nodes, 0 is a matrix of zeroes and $I_R$ is the $|V_R| \times |V_R|$ sub-matrix representing absorption at a consumer node.

Starting from $s^*$ , the diffusion of a contaminated unit through the supply network is fully determined by the Markov transmission matrix $P$. The process ends when an absorbing node $o \in V_R$ is reached, generating a list of directed edges connecting $s^*$ and $o$, the network path $\gamma_{s^*o}$. After departing the supply network at $o$, the contaminated unit is consumed. $A$ set of $K$ individuals consuming contaminated units will report or observe illness. We label the node linked to observation $k$ by $o_k$ , resulting in a list of $K$ observations $\Theta = (o_1,\,\ldots,\,o_K )$. These observations will be linked to the network at the unique set of nodes $o \in \Theta \subseteq V_R$ , such that $|\Theta| \leq K$ . An important implication of the ``pinball'' assumption is that the transmission paths $\gamma_{so_k}$ through the supply network, and thus the observations $o_k$ , are mutually independent.

The final step in developing the transmission model involves connecting the stochastic process with the physical quantities defined in the network model. The volumes shipped from $i$ to $j$ can be seen as a proxy for the conditional probability that a contaminated item is sent along that direction. We therefore define the transition probabilities $p_{ij}$ as the proportion of volume-flux sent from $i$ to $j$,
\[
p_{ij}=\frac{w_{ij}}{\displaystyle\sum_jw_{ij}}\in[0,1].
\]
\subsection{Source Estimator: Bayesian Inference}

Our goal is to find the most ``probable'' source $s^* \in V_Q$ based on the list of observations $\Theta$. We introduce a Bayesian formulation for the probability that a feasible source node $s$ is the true source $s^*$, given the observations and the prior distribution over $s^*$:

\begin{equation}
P(s^*=s|\Theta) = \frac{P(s^*=s)P(\Theta|s^*=s)}{P(\Theta)}
\label{eq:01}
\end{equation}

To identify the source, we adopt a maximum probability of detection approach, designing an estimator $\hat{s}$ that selects the feasible source node $s$ that maximizes the probability $P(s^*=s|\Theta) $, i.e.

\begin{equation}
\hat{s} = \underset{s\in\Omega}{\mbox{arg max}}P(s^*=s)P(\Theta|s^*=s)
\label{eq:02}
\end{equation}

where $s \in\Omega$ is the set of feasible source nodes. Here we have observed that only a subset of nodes $\Omega \subseteq V_Q$ will be feasible source candidates: the set of nodes in $V_Q$ that share at least one path through the network to all contaminated nodes $o_k \in\Omega$. Unless any prior information regarding the source location is available\footnote{Prior information, or information external to the network, may be available regarding the location of the source. This may come from known risk factors, e.g. extreme weather events or sighting of feral wild animals, that increase the prevalence of contamination. It may also come from expert opinion.} we assume the prior distribution over $s^*$ is uniform, i.e. $P(s^*=s)= 1/|\Omega|$ for all nodes $s \in\Omega$ , making the estimator $\hat{s}$ the maximum likelihood estimator.

The main challenge in solving (\ref{eq:02}) is estimating the likelihood $P(\Theta|s^*=s)$. The probability of observing illnesses at the locations in $\Theta$ from a contamination originated at $s$ will depend on the paths taken through $G$ from s to all observations $o_k \in\Theta$ . However, there are multiple possible paths $\gamma_{so_k}$ from $s$ to each observation node $o_k \in \Theta$. The exact probability that s is the source is equal to the total probability over every permutation of paths for which s is the source. We now introduce a few definitions that allow us to write the source likelihood defined over all permutations:

\begin{itemize}
\item $\Gamma_{so_k}=$ The set of all paths $\gamma^{\{n\}}_{so_k}$ through $G$ from $s$ to $o,\left\{\gamma_{so_k}^{\{1\}},\ldots,\gamma_{so_k}^{\{N\}}\right\}$
\item $\pi_s =$ An \textit{s-cascade}, or a specific permutation of the $K$ paths connecting feasible source $s$ to each observation location $o_k \in\Theta$ . Formally, $\pi_s$ is an element of the Cartesian product over the sets of paths $\left\{\Gamma_{so_k}\right\}_{o_k\in\Theta}$, i.e.,\\ $\left(\gamma_{so_1},\ldots,\gamma_{so_K}\right)\in\Gamma_{so_1}\times\cdots\times\Gamma_{so_K}$

\item $\Pi_s =$ The set of all $s$-cascades, or all permutations of paths from $s$ to each $o_k \in\Theta$ , i.e.,
\begin{equation}
\Pi_s = \Gamma_{so_k}\times\cdots\times\Gamma_{so_k}=\left\{(\gamma_{so_1},\ldots,\gamma_{so_K}):\gamma_{so_k}\in\Gamma_{so_k}\right\}
\label{eq:03}
\end{equation}
\end{itemize}

With these definitions, the source likelihood can be written as the total probability of all permutations of paths where $s$ is the source of the cascade,

\begin{equation}
P(\Theta|s^*=s)=\sum_{\pi_s\in\Pi_s}P(\pi_s|s)P(\Theta|s,\pi_s)=\sum_{\pi_s\in\Pi_s}P(\pi_s|s),
\label{eq:04}
\end{equation}

where the term $P(\Theta|s,\pi_s)$ is equal to 1, since by definition, the observation locations $o_k\in\Theta$ are the endpoints of the paths $\gamma_{so_k}\in\pi_s$.

Solving equation (\ref{eq:04}) amounts to finding the total probability over all $s$-cascades $\pi_s \in\Pi_s$ . The probability of an individual $s$-cascade $P(\pi_s|s)$ can be expanded in terms of the constituent paths $\gamma_{so_k} \in\pi_s$ and transition probabilities $p_{ij}$ between each adjacent node pairs $(i, j)\in\gamma_{so_k}$ as
\begin{eqnarray}
P(\pi_s|s) & = & P(\gamma_{so_1},\ldots,\gamma_{so_K}|s)\nonumber\\
\label{eq:05}
 & = & \displaystyle\prod_{\gamma_{so_k}\in\pi_s}P(\gamma_{so_k}|s)\\
  & = & \displaystyle\prod_{\gamma_{so_k}\in\pi_s}\prod_{(i,j)\in\gamma_{so_k}}p_{ij}\nonumber
\end{eqnarray}

where the second equality follows from the independence of paths to each observation $o_k$ and the third equality follows from the total probability associated with path $\gamma_{so_k}$. The likelihood can then be found in terms of the
transition probabilities as

\begin{equation}
P(\Theta|s^*=s) = \sum_{\pi_s\in\Pi_s}\prod_{\gamma_{so_k}\in\pi_s}\prod_{(i,j)\in\gamma_{so_k}}p_{ij}.
\label{eq:06}
\end{equation}

An illustration evaluating equation (\ref{eq:06}) on a small network and outbreak is provided in Section~\ref{sec:toyEx}.\\

Evaluating the likelihood by solving equation (\ref{eq:06}) explicitly requires enumerating all paths $\gamma_{so_k}$ contained in each \textit{s-cascade} $\pi_s$ , over all cascades $\Pi_s$ , which becomes combinatorially difficult for large networks given even very few illness observations. Existing methods have dealt with this complexity by assuming the contamination travels along a single \textit{s-cascade}: the set of shortest paths \cite{24,25} or shortest, highest probability paths \cite{26,27}. In the following we introduce an alternate representation of the likelihood $P(\Theta|s^*=s) $ that allows us to develop a simple algebraic expression that is probabilistically equivalent to equation (\ref{eq:06}), from which we can tractably compute the total probability over all \textit{s-cascades}.

We begin by showing that equation (\ref{eq:04}) can be rearranged from an expression that enumerates over each \textit{s-cascade} $\pi_s\in\Pi_s$ to one that enumerates over each observation $o_k\in\Theta$. Starting from the right-hand side of (\ref{eq:04}) we have,
\begin{equation*}
\sum_{\pi_s\in\Pi_s}P(\pi_s|s) = P(\Pi_s|s) = P(\Gamma_{so_1}\times\cdots\times\Gamma_{so_K}|s),
\end{equation*}

by total probability and the definition of $\Pi_s$ . Then,

\begin{equation*}
P(\Gamma_{so_1}\times\cdots\times\Gamma_{so_K}|s) = P\left(\prod_{o_k\in\Theta}\Gamma_{so_k}|s\right) = \prod_{o_k\in\Theta}P\left(\Gamma_{so_k}|s\right).
\end{equation*}

where the last equality follows from the independence of observations $o_k \in\Theta$. Therefore,

\begin{equation}
\sum_{\pi_s\in\Pi_s}P(\pi_s|s) = \prod_{o_k\in\Theta}P\left(\Gamma_{so_k}|s\right).
\label{eq:07}
\end{equation}

The term $P\left(\Gamma_{so_k}|s\right)$ represents the total probability of reaching location $o_k$ from starting point $s$ along all possible paths $\gamma_{so_k}^{\{n\}}\in\Gamma_{so_k}$. Let us denote this probability as  $P\left(\Gamma_{so_k}|s\right)=a_{so_k}$, so that we are interested in evaluating

\begin{equation}
P(\Theta|s^*=s) = \prod_{o_k\in\Theta}a_{so_k}.
\label{eq:08}
\end{equation}

To compute $a_{so_k}$ we could sum over the probability of all individual paths $\left\{\gamma_{so_k}^{\{1\}},\ldots,\gamma_{so_k}^{\{N\}}\right\}\in\Gamma_{so_k}$, but this again requires enumerating all possible paths between $s$ and $o_k$, which is as combinatorially difficult as evaluating (\ref{eq:06}).\\

An alternative representation involves recognizing $a_{so_k}$ as the \textit{absorbing probability} for a Markov chain, or the probability that a contaminated item starting at $s$ gets ``captured'' at $o_k$ . The absorbing probability can be written as \cite{31}:

\begin{equation}
a_{so_k} = \sum_{n=0}^{\infty}\sum_{l\in V_Q}p_{sl}^{(n)}p_{lo_k},
\label{eq:09}
\end{equation}

where $p_{sl}^{(n)}$ denotes the probability of transitioning from transient node $s$ to transient node $l \in V_Q$ in exactly $n$ steps, and $p_{lo_k}$ denotes the probability of transitioning from $l$ to absorbing node $o_k$ (in one step). Equation (\ref{eq:09}) represents the probability of starting at $s$ and being absorbed at $o_k$ in one or more steps -- that is, over paths of any length\footnote{This is a general formulation that allows paths of different lengths to the observations. If no supply network edges exist between nodes within a stage or to a previous stage, there will be no cycles in $G$ and $n$ will be bounded.}. The probability of being absorbed in a single step is equal to $p_{so_k}$ . If this does not happen, the contamination may move either to another absorbing state (in which case it never reaches $o_k$ ), or to transient state $l$. In fact, it may move among the transient states for any number of transitions before landing at $l$, which occurs after $n$ steps with probability $p_{sl}^{(n)}$. From $l$ it then has probability $p_{lo_k}$ of going to $o_k$.

The probability $p_{sl}^{(n)}$ that the contamination travels from $s$ to $l$ in exactly $n$ steps is found as the $(s,l)^{\mbox{th}}$ element of $sl$ the transition-state matrix $P_Q$ raised to the $n$th power. Therefore, we can write equation (\ref{eq:09}) in matrix form as \cite{31},

\begin{equation}
A=\sum_{n=0}^{\infty}P^n_QP_R,
\label{eq:10}
\end{equation}

where we have also recognized $p_{lo_k}$ as the $(l,o)^{\mbox{th}}$ element of the absorbing-state matrix $P_R$. Here we distinguish $o$ and $o_k$, since $o$ describes the unique node in $V_R$ corresponding to the observation $o_k$ and therefore $(l,o)$ points to a specific entry in $P_R$. Summing the geometric series, equation (\ref{eq:10}) can be expressed in closed-form as,

\begin{equation}
A=(I-P_Q)^{-1}P_R,
\label{eq:11}
\end{equation}

which is well-defined because for any absorbing Markov chain, $I-P_Q$ will have an inverse \cite{31}.

Combining equations (\ref{eq:08}) and (\ref{eq:11}) we can fully define the likelihood over all observations,

\begin{equation}
P(\Theta|s^*=s) = \prod_{o_k\in\Theta}\left[(I-P_Q)^{-1}P_R\right]_{so_k}.
\label{eq:12}
\end{equation}

Evaluating this equation requires only a single operation to compute the matrix $A$.

We can now see the full advantage of the relation derived in equation (\ref{eq:07}): whereas the left-hand term requires enumerating all possible paths $\gamma_{so_k}$ between $s$ and each observation $o_k$, by rearranging to order over the observations, the right-hand term can be evaluated through a single algebraic computation.

Equation (\ref{eq:12}) can be used in the maximum likelihood source estimator in equation (\ref{eq:02}) to select the source node that maximizes the posterior probability $P(\Theta|s^*=s)$ over all possible sources $s \in\Omega$,
\begin{equation}
\hat{s} = \underset{s\in\Omega}{\mbox{arg max}}\prod_{o_k\in\Theta}\left[(I-P_Q)^{-1}P_R\right]_{so_k}.
\label{eq:13}
\end{equation}

We can also construct a posterior probability for each feasible source $s \in\Omega$,
\begin{equation}
P(s^*=s|\Theta) = \frac{1}{c}P(s^*=s)P(\Theta|s^*=s),
\label{eq:14}
\end{equation}

for some normalizing constant $c$, forming a probability distribution over the set $s\in\Omega$, which can be used to identify a set of the most probable sources.

It is important to note that by the formulation in equation (\ref{eq:09}), $a_{so_k}$ represents the total probability of reaching location $o_k$ from starting point $s$, considering all possible paths of all possible lengths. Given the transmission model of Section~\ref{sec:transmissionModel}, which assumes observations are independent, the likelihood in \eqref{eq:12} thus represents the exact total probability of all observations resulting from $s$.

This source detection approach relies on the independent observation assumption. The implication of this assumption is that contamination trajectories or \textit{s-cascades} containing paths with shared edges are not assigned higher probabilistic weighting during inference of the source. This assumption may introduce some error in situations where contaminated items have traveled in the same batch through early legs of their journey through the supply network, for example being shipped together from producer to distributor before being divided into separate pallets. Nonetheless, the condition of independence between observations can reasonably be expected to be validated in practice, since it is possible for food items from the same contaminated batch to depart from the source in separate (and independent) trucks; indeed, for large contamination incidents where the contaminated quantity will be larger than what fits in one truck, this is necessarily the case. We therefore expect the error caused by failing to consider shared pathways to be of second order and that our solution is a good approximation of the maximum likelihood source estimator.

\subsection{\label{sec:toyEx}Illustration}

\begin{figure}[h]
\includegraphics[width=.50\textwidth]{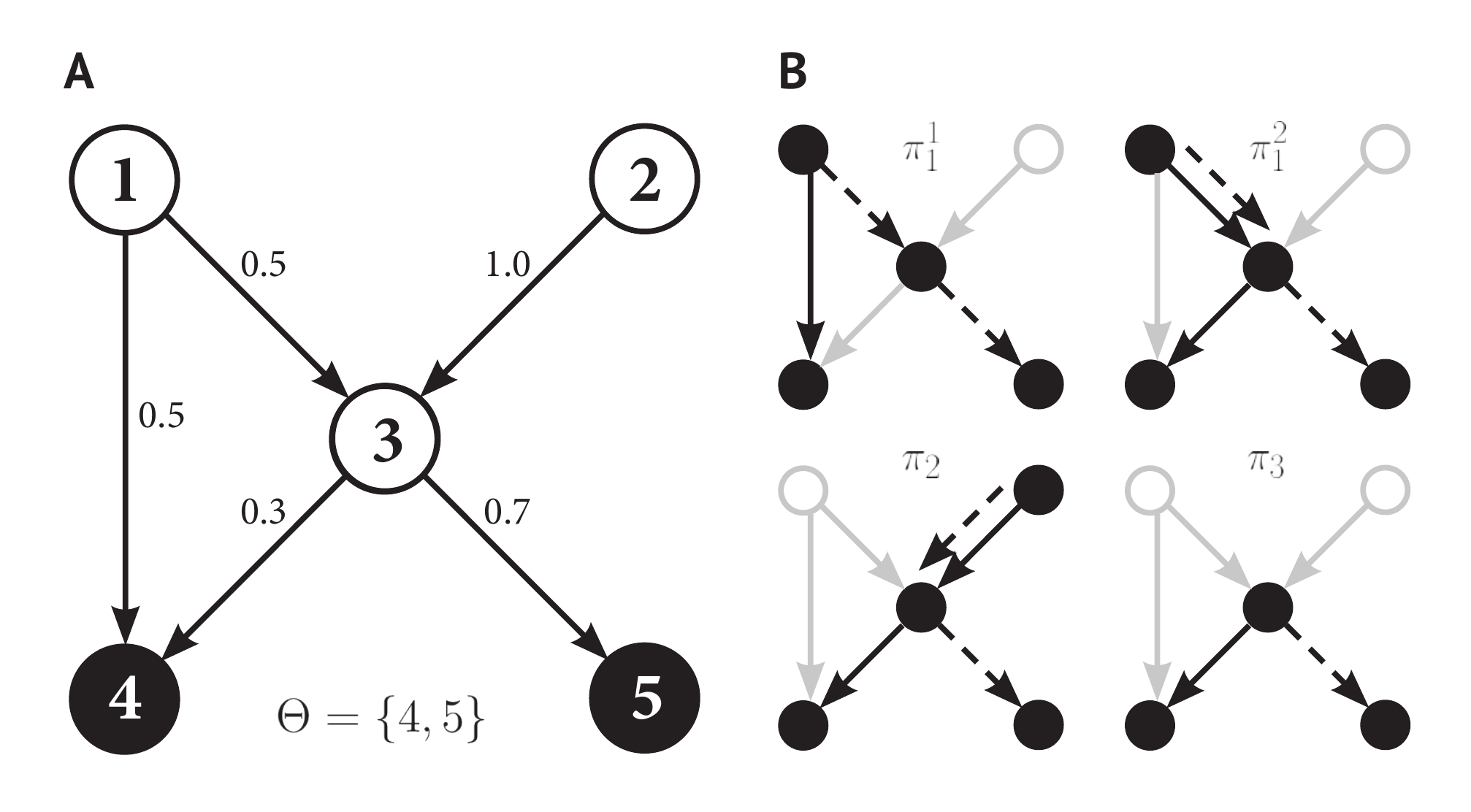}
\caption{(A) An example outbreak over a simple network with two producer nodes, two retailer nodes, and one intermediary node. Edges are annotated with transition probabilities, and the shaded nodes represent observations of contamination. (B) The set of all cascades that could have possibly induced the outbreak in A. Solid and dashed lines show paths of contamination to nodes 4 and 5, respectively. Shaded nodes are those through which contamination has passed. $\pi_1^1$ is the higher-probability cascade in $\Pi_1$ (the set of all possible cascades starting from node 1), and $\pi_1^2$ is the lower-probability cascade in $\Pi_1$. $\pi_2$ and $\pi_3$ are the only cascades in $\Pi_2$ and $\Pi_3$, respectively.}
\label{fig:toyEx}
\end{figure}

To demonstrate our method of calculating source probabilities -- as well as the benefit of considering all possible paths -- we now present an example outbreak over a small food supply network.

We take $G$ to be the network pictured in Fig.~\ref{fig:toyEx}. Additionally, we consider a two-observation outbreak, consisting of one illness at node 4 and another illness at node 5. 

We note that four cascades could possibly account for this outbreak -- they are pictured in Fig.~\ref{fig:toyEx}B -- and we use this information to tractably calculate a source estimator through equation (\ref{eq:06}). 

Through equation (\ref{eq:05}), we evaluate the probabilities of each of these cascades to be:
$$ P(\pi_1^1 | s^* = 1) = 0.5 \times 0.5 \times 0.7 = 0.175 $$
$$ P(\pi_1^2 | s^* = 1) = 0.5 \times 0.3 \times 0.5 \times 0.7 = 0.0525 $$
$$ P(\pi_2 | s^* = 2) = 1.0 \times 0.3 \times 0.7 = 0.21 $$
$$ P(\pi_3 | s^* = 3) = 0.3 \times 0.7 = 0.21 $$
Which gives us, through equation \ref{eq:06}:
$$ P(\Theta | s^* = 1) = 0.175 + 0.0525 = 0.2275 $$
$$ P(\Theta | s^* = 2) = 0.21 $$
$$ P(\Theta | s^* = 3) = 0.21 $$

and thereby leads us to select node 1 as our most likely source. 

Expectedly, taking 
$$P_Q = \begin{bmatrix}
0 & 0 & 0.5 \\ 
0 & 0 & 1 \\ 
0 & 0 & 0
\end{bmatrix}$$

$$P_R = \begin{bmatrix}
0.5 & 0\\ 
0 & 0\\ 
0.3 &0.7 
\end{bmatrix}$$
 and evaluating the conditional outbreak probabilities via the matrix formulation posed in equation (\ref{eq:12}) yields the same values, and therefore estimator.
 
Note, however, that our answer hinges on our method having accounted for all possible paths: considering only the most probable cascades from each source would have resulted in selection of node 2 or node 3 as the most probable source despite the fact that node 1 delivers more total volume to the sites of illness observation.


\section{\label{sec:Evaluation}Evaluation}

In Section~\ref{sec:Model} we derive the ML estimator for the source of an outbreak of foodborne disease given an underlying network structure. In this section we demonstrate the performance benefits of the probabilistically exact source estimator in application to different network topologies. First we apply the method to stylized network models of the food supply and simulated outbreaks of contamination. This allows us to evaluate the performance of our ML source estimator and its robustness to differences in network structure in an idealized setting. Applying the method to stylized networks is equivalent to assuming that the exact, complete network and outbreak data is available for source location. In practice, food supply networks are never exactly known, and illness data is imperfect, especially during an unfolding outbreak when data is emerging. In order to evaluate the robustness of our conclusions in these non-ideal settings we also apply our method to illness data from a real outbreak, using an estimated model of the relevant food supply network structure. We introduce a simple and replicable way to model the network based on available statistical data and basic understanding of food supply logistics. Because the estimated network is key to the implementation of the method, this application represents an integrated evaluation of the ML source estimator combined with the network modeling methodology.

\subsection{\label{sec:stylizedNets}Evaluation on stylized networks}

\subsubsection{Stylized network structures}

We first evaluate our method on stylized models of food supply networks. We choose for application one characteristic food supply structure: a layered, directed network consisting of four layers of supply, for which nodes in each layer trade expressly with nodes in the subsequent layer; this is the structure exhibited by the network in Fig.~\ref{fig:01}. Formally, this is a network of the form $G ={V,E}$ with four layers $V =\{V_1,V_2,V_3,V_4\}$, and directed edges of the form $(i,j)\in E$ for $i\in V_{n}$, $ j\in V_{n+1}$, $n=1,2,3$. We consider two probabilistically different network topologies based on this characteristic structure, which we use to evaluate the source detection methods. On one extreme is a structure for which a small percentage of edges carry the majority of the probability weight. The dominant probabilities will capture a large fraction of the product flowing through the network; therefore we call this the \textit{dominant paths} network. On the other extreme is a structure for which each edge is probabilistically equivalent to every other. Paths through this network will also be probabilistically equivalent and no path will capture more of the flow through the network than any other; we therefore call this the \textit{non-dominant paths} network. We purposefully choose these two topologies that, in addition to being relevant examples of food supply network structure, enable us to evaluate the robustness of source detection methods to differences in path probabilities.

The differing probabilities along the edges are determined by how nodes are connected. To create the \textit{dominant} paths network, we use a connection scheme for which there is high variance in the degree $D$, or the number of edges that any given node is connected to. To achieve a wide distribution we sample the degree d of each node according to a geometric distribution, i.e. $D \sim geom(1/\mu_D )$, where $\mu_D$ is the pre-defined average degree across all nodes. After assigning each node an in- and out-degree, representing the number of incoming and outgoing edges, we randomly pair edges to nodes in adjacent layers according to the Network Configuration Model \cite{32}. We then divide the probabilities equally across all outgoing edges from each node, i.e. $p_{ij} = 1/d = 1/|j|$. Edges coming from highly connected nodes will therefore carry lower probability weight. This can be interpreted as a dichotomy between supply network actors with a large number of small trading partners and actors with a small number of large trading partners. For the non-dominant paths network, we use a deterministic connection scheme such that each node is connected to exactly the same number of nodes as any other node within the layer. The in- and out- degrees are thus the same for all nodes in a layer, equal to $\mu_D$ , and the probabilities are equal for all edges, i.e. $p_{ij} =1/\mu_D$.

\begin{figure}[b]
\includegraphics[width=0.22\textwidth]{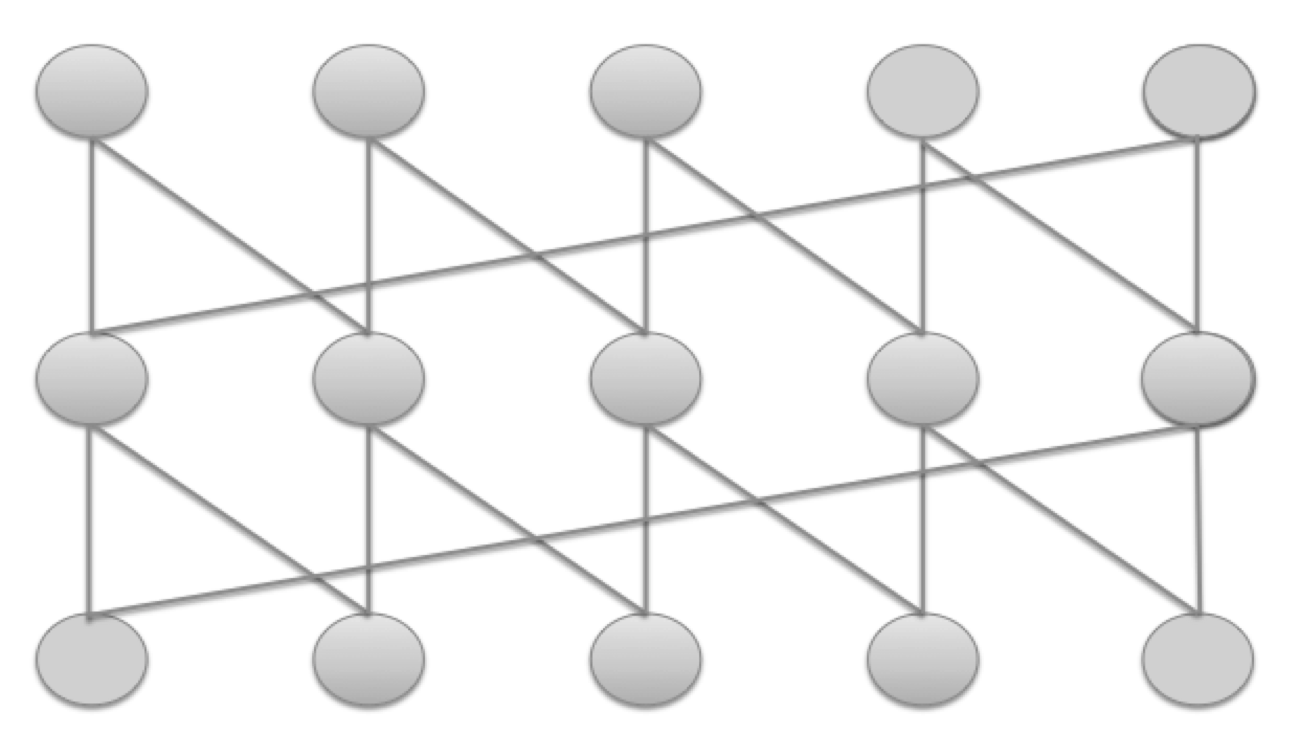}
 \includegraphics[width=0.22\textwidth]{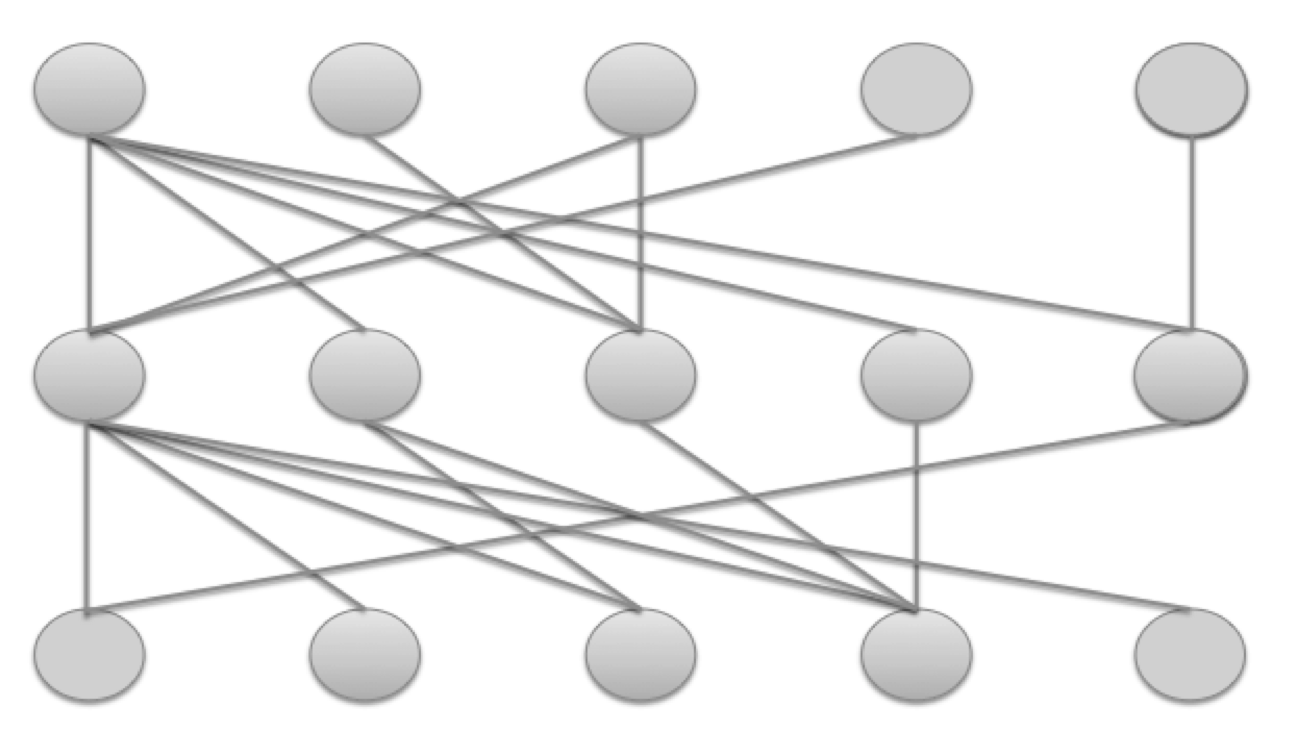}
 \begin{subfigure}[t]{0.2\textwidth}
\caption*{(a) Non-Dominant Paths}
\end{subfigure}
 \begin{subfigure}[t]{0.2\textwidth}
\caption*{(b) Dominant Paths}
\end{subfigure}
\caption{\label{fig:stylizedNets}Toy example illustrating the structure of the stylized networks analyzed in Section~\ref{sec:stylizedNets}. Both networks have the same number of nodes, links, and average degree, but the degree distributions vary:\\
(a) the Non-Dominant Paths Network is generated according to a deterministic degree distribution, while
(b) the Dominant Paths Network is generated according to a geometric degree distribution.}
\end{figure}

For the simulations presented in this section we fix 25 nodes per layer, and we choose an average degree of $\mu_D = 4$. Many different network structures with these parameters can be created according to the network generation process described both for dominant and non-dominant paths connection schemes; we illustrate stylized versions of these structures in Fig.~\ref{fig:stylizedNets}.

\subsubsection{Benchmarks}

Throughout this section we will compare the performance of our method to the \textit{Effective Distance} method for source identification proposed in \cite{26,27} and to a benchmark indicating the accuracy of random guessing, which we call the \textit{Network Baseline}. As discussed in Section~\ref{sec:ProblemFraming}, the Effective Distance method is the state-of-the-art approach for source detection on weighted networks, and the method most relevant in the context of the foodborne disease source detection. The other methods reviewed in Section~\ref{sec:ProblemFraming} are either not implementable or impractical due to shortest path assumptions. In the following we describe the Effective Distance method in terms of the notation introduced in Section~\ref{sec:Model}.\\

\noindent\textbf{\textit{Effective Distance}}

The Effective Distance method for source identification \cite{26,27} is based on the concept that the trajectory of a particle diffusing through a network will primarily follow the single shortest, highest probability path to any other node. The true source $s^*$ of an outbreak should therefore be the node that exhibits the set of shortest, highest probability paths to the outbreak node set $o \in \Theta$.

Based on this logic, the authors introduce a metric for the Effective Distance $d_{eff} (i, j)$ between two connected
nodes $i$ and $j$, defined such that the likelier the connection, the shorter the Effective Distance. This is given as
\begin{equation}
d_{eff} (i, j) = 1-\log p_{ij},
\label{eq:15}
\end{equation}

where $p_{ij}$ is the probability of transiting from $i$ to $j$ as defined in Section~\ref{sec:Model}. The effective length of a given path $\gamma_{so}$ between nodes $s$ and $o$ is then defined to be the sum total of the Effective Distances of each edge $(i, j )\in \gamma_{so}$. As discussed, the concept of \cite{26,27} is to focus on the single shortest Effective Distance path over all possible paths $\gamma_{so} \in\Gamma_{so}$ from $s$ to $o$. Therefore the Effective Distance between $s$ and $o$ is defined as,
\begin{eqnarray}
D_{eff} (s,o) & = & \min_{\gamma_{so}\in\Gamma_{so}}  \sum_{(i,j)\in\gamma_{so}}1-\log p_{ij}\nonumber\\ 
&=& \min_{\gamma_{so}\in\Gamma_{so}}[|\gamma_{so}|-\log P(\gamma_{so}|s)]
\label{eq:16}
\end{eqnarray}

The inclusion of the logarithmic term allows path probabilities to be determined by addition instead of multiplication, and the shortest Effective Distance path can be identified with Dijkstra's shortest path algorithm using $d_{eff} (i, j)$ for the length of each edge. The Effective Distance of a path therefore results from a multifactorial objective function that penalizes long path lengths while rewarding high path probabilities.

The measure of Effective Distance is used in source detection in the following way. For each feasible source node
$s \in\Omega$ the set $\pi_s^{D_{eff}}$ of shortest Effective Distance paths to all unique observation nodes $o \in \Theta$ is identified. Each $s$ observation node $o \in \Theta$ is counted only once for disease incidence, so $\pi_s^{D_{eff}}$ represents an \textit{s-cascade} with no $s$ repeated paths. An estimator $\hat{s}_{D_{eff}}$ then chooses the source node that minimizes the mean $\hat{\mu}_\Theta (D_{eff} (s,o))$ and variance $\hat{\sigma}^2_\Theta (D_{eff} (s, o))$ of the Effective Distances of the paths in $\pi_S^{D_{eff}}$, according to the objective
\begin{equation}
\hat{s}_{D_{eff}} = \underset{s\in\Omega}{\mbox{arg min}}\sqrt{\hat{\mu}^2_\Theta(D_{eff}(s,o))+\hat{\sigma}_\Theta^2(D_{eff}(s,o))}.
\label{17}
\end{equation}

This estimator is in effect identifying the source with the set of shortest Effective Distances to the observation node set $\Theta$, offset by a variance term that discounts extreme values.

It is important to note that when all paths $\gamma_{so} \in\Gamma_{so}$ from $s$ to $o$ are the same length, the Effective Distance path identified by (\ref{eq:16}) will be the highest probability path from $s$ to $o$. In these cases, the Effective Distance estimator $\hat{s}_{D_{eff}}$ is performing source identification by considering the single set of highest probability paths to the observation nodes $o\in \Theta$, which is the \textit{s-cascade} $\displaystyle \pi_s^{D_{eff}} = \max_{\pi_s\in\Pi_s}P(\pi_s| s)$.\\
\\

\noindent\textbf{\textit{Network Baseline}}

We also compare results to the Network Baseline, a benchmark that is equivalent to guessing at random between all feasible sources $s \in\Omega$. Comparisons to this benchmark demonstrate that the accuracy of source identification is not determined by the \textit{number} of feasible sources.

\subsubsection{Simulation setting}

For each network structure, we generate outbreaks by using each node in layer $V_1$ as the source. A Monte Carlo simulation model determines the trajectories of contamination through the supply chain, leading to observations of illness at the set of node locations $o_k \in\Theta$. At a snapshot in the outbreak's progression, the source detection methods are applied and all feasible sources are rank-ordered according to their probability values or Effective Distances. This process is repeated at increasing intervals in each outbreak's progression, generating a series of rankings as a function of the number of cases. We run 50 simulations per $V_1$ node, for a total of 1000 outbreak simulations. To assess the traceback performance for each network structure, the cumulative results of source detection are quantified according to \textit{Simulation Accuracy}, which measures the percentage of times the true source is accurately identified across all simulations.

\subsubsection{Results}

\begin{figure}[b]
\includegraphics[width=0.23\textwidth]{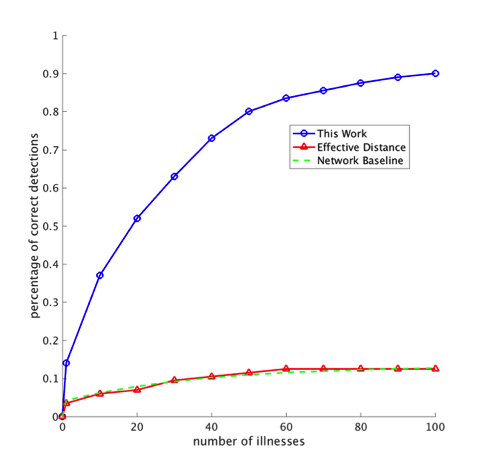}
 \includegraphics[width=0.23\textwidth]{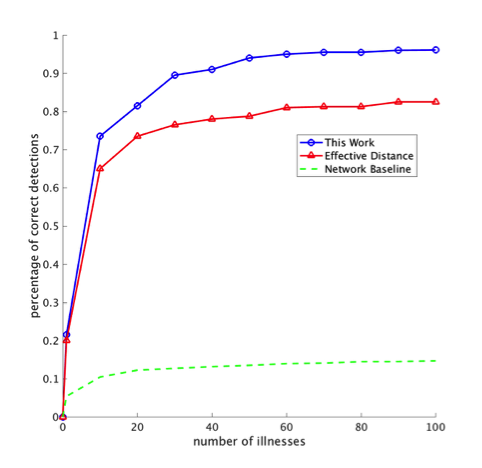}
 \begin{subfigure}[t]{0.2\textwidth}
\caption*{(a) Accuracy: Non-Dominant Paths}
\end{subfigure}
 \begin{subfigure}[t]{0.2\textwidth}
\caption*{(b) Accuracy:\\Dominant Paths}
\end{subfigure}
\caption{\label{fig:stylizedNetsResults}Simulation Accuracy for the networks introduced in Section~\ref{sec:stylizedNets} for our method for source detection (blue, $\bigodot$), the Effective Distance method for source detection (red, $\Delta$), and the Network Baseline (green, - -) as a function of the number of illnesses reported.}
\end{figure}

Fig.~\ref{fig:stylizedNetsResults} demonstrates results for Simulation Accuracy with our source detection method, the Effective Distance method, and the Network Baseline as a function of the number of cases for the \textit{non-dominant paths network} (~\ref{fig:stylizedNetsResults}a, left) and the dominant paths network (~\ref{fig:stylizedNetsResults}b, right). The Network Baseline or random guessing is included to demonstrate that results are not due to a lack of feasible sources; if there would be a lack of feasible sources, the network baseline would increase in accuracy very quickly, which is not the case.

For both networks our method performs well (see Figs. ~\ref{fig:stylizedNetsResults}a and~\ref{fig:stylizedNetsResults}b) and follows expected properties, increasing in accuracy with data on the number of illness reports. We can make good inferences about the source location after only a limited number of illnesses have been reported, and very accurate inferences if we wait a bit longer. The accuracy of source identification is faster and more accurate for the \textit{dominant paths network}; this is as expected since high probability edges will dominate the paths traveled by contaminated product as well as the calculation of the source identification likelihood. Despite the lack of dominant path probabilities, Simulation Accuracy is also high for the \textit{non-dominant paths network}. What is happening is that when the probabilities of all paths are equal, our method reduces to calculating the number of possible \textit{s-cascades} between a source and set of observations; this can be seen by replacing the $p_{ij}$ in equation (\ref{eq:06}) with a constant. This effectively turns our source estimator into a centrality-based method that chooses the source that connects to the observations across the greatest number of paths.

The comparison of Simulation Accuracy results for our source estimator and the Effective Distance method make apparent the benefit of considering all paths in estimating the source rather than selecting only the set of highest probability paths. On the \textit{non-dominant paths network}, the Effective Distance method cannot compete; because all paths appear the same, the method chooses one at random, and as a result performs identically to the Network Baseline. Still, this network is a stylized and extreme case; most real-world food supply networks will exhibit some degree of heterogeneity in path probabilities. On the \textit{dominant paths network}, which exhibits significant heterogeneity in path probabilities, the Effective Distance method performs much better. This is as expected: for each feasible source the method considers the highest probability \textit{s-cascade}; when path probabilities vary greatly, this will often be the actual set of paths traveled by the contamination. However because the contamination does not always travel along the highest probability \textit{s-cascade}, by accounting for all possible \textit{s-cascades}, our method performs better by a substantial margin of around 10\%.

Since these two networks represent extremes in the way probabilities might be distributed across a food supply network, the results presented here suggest that source identification in the context of foodborne disease is robustly and substantially more accurate when the total probability across all possible paths between a feasible source and the observation set is considered.

\subsection{Application: 2011 EHEC O104:H4 Outbreak}

In this section we evaluate our method in application to the Shiga toxin-producing Escherichia coli (EHEC) O104:H4 outbreak in Germany in 2011, which affected over 4,000 people with EHEC gastroenteritis or severe hemolytic uremic syndrome (HUS). Beyond its being a widely-known, high-profile event, we choose this case for application for the following reasons. First, it is a relevant illustration of the type of large-scale distributed outbreak that motivated the design of our method. Second, data informing an estimate of the underlying food supply network structure is available for Germany, allowing us to implement our source identification methodology. Third, it was conclusively solved, meaning traced to a specific origin location. This permits us to verify the accuracy of the results of our method. Lastly, it allows comparison to published results on the application of the existing best-in- class method \cite{27} to the same case. This enables us to complement the evaluation of the comparative effectiveness of the source detection methods on stylized data presented in Section~\ref{sec:stylizedNets} with a practical evaluation in application to real data.

\subsubsection{EHEC 2011 outbreak background and investigation timeline}

\begin{figure}[b]
\begin{center}
\includegraphics[width=0.5\textwidth]{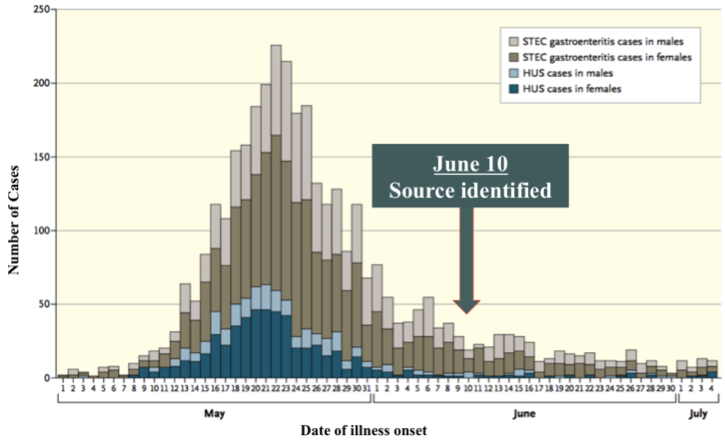}
\end{center}
\caption{\label{fig:EHECcurve}The epidemiological curve of the EHEC O104:H4 outbreak, depicting the number of cases by date of illness onset. The first case of illness was reported on May 4 and the last case on July 4, 2011. The source of the outbreak, contaminated sprouts from an organic farm in Bienenb\"{u}ttel, district Uelzen, was confirmed on June 10, 2011. Image source: \cite{1}.}
\end{figure}

Fig.~\ref{fig:EHECcurve} depicts the epidemic curve of the outbreak. The first confirmed illness case began on May 1$^{st}$, marking the beginning of Week 1. The case count grew dramatically starting on May 8$^{th}$, at the beginning of Week 2. The outbreak peaked on May 21 and 22, between Week 3 and Week 4, and the majority of cases had been reported by the end of Week 5. The last illness associated with the outbreak was reported on July 4, at the end of Week 9 going into Week 10, but the outbreak was declared over 3 weeks later, on July 26$^{th}$. \cite{1,2}\\

Epidemiological interviews with patients in the first outbreak clusters and subsequent analyses suggested that (i) raw vegetables were the source of the infection, and (ii) these vegetables were primarily consumed at restaurants \cite{2,33}. By Week 6 investigators had narrowed the search down to contaminated sprouts and by the end of that week, on June 10, confirmed the origin of the outbreak as a small organic farm in Bienenbüttel, in the district Uelzen in Northern Germany. At this time a public service announcement was issued advising the public to avoid consuming raw sprouts, and the producer in Bienenbüttel was temporary closed. Unfortunately, by this point the contaminated product had made its way through the supply chain and the majority of illnesses had occurred, as is made clear in Fig.~\ref{fig:EHECcurve}.

The outbreak was one of the most virulent and deadly in recent history, leading to 4,321 total outbreak cases, including 3,469 cases of EHEC gastroenteritis, 852 cases of HUS, and 54 reported deaths \cite{1,2}. The majority of these illnesses (88\%) and all of the deaths were in Germany, and in particular northern Germany, but cases were distributed across 16 countries including a few reports in North America \cite{1,34}.

\subsubsection{Case data}

The illness case data comes from the Robert Koch Institute (RKI), Germany's national public health authority, by way of the ServStat tool \cite{35}. We query this tool for data on all cases of E.coli reported in Germany during the dates May 1 to July 4, 2011, corresponding to outbreak Weeks 1 through 9. This includes all cases of any strain of E.coli contamination reported in Germany during this time period, including cases unrelated to the EHEC O104:H4 outbreak. Since the number of cases known to be linked to this outbreak exceed the routine baseline of E.coli contamination by a factor of 70 times during this time period in similar years \cite{1}, these unrelated cases represent a minor source of noise in the data.\\

Cases are reported in association with the German ``Landkreise'' or administrative district where the patient resides. There are a total of 402 districts in Germany. We use only this data and do not consider cases reported outside of Germany.

\subsubsection{Network model}

The source identification method described in Section~\ref{sec:Model} assumes a model of the underlying food supply network structure. Because exact, fine-grained data on the supply network of food commodities are not available, we develop a model of the network based on publicly available data and a practical understanding of food supply chain logistics. We focus this model on the supply of vegetables in Germany because (i) raw vegetables were suspected as the source of infection early in the investigation and (ii) most of the reported cases of illness were inside Germany.\\

As basis for the network model we use a probability matrix $P_V$ for vegetables in Germany. The elements of this
matrix $p_{ij}$ represent the probability of vegetables being transported from region $i$ to $j$, such that $\displaystyle \sum_jp_{ij} =1$. The vegetable data originates from a meta-model of the food supply in Germany differentiating between the 402 German administrative districts and 50 import countries \cite{36}. To estimate the data the meta-model uses gravity models calibrated with transport matrix data from the German Infrastructure Master Plan \cite{37}. The gravity models are based on available statistics from agricultural production, consumption, and sectorial interrelations. For more details on the modeling approach and data sources see \cite{36}. This work represents, to the best of our knowledge, the only model of the German food supply that is also calibrated and is therefore the best available data on which to base our network model.\\

Many of the reported cases of illness occurred in clusters linked to restaurants \cite{2}. Restaurants primarily get their deliveries from wholesalers. We assume that the path of vegetables via wholesalers has three edges: one from the farmer or producer to the wholesaler specialized on vegetables, one from the specialized wholesaler to the wholesaler doing the distribution, and one from the distribution wholesaler to the restaurants. While this is an approximation, interviews with practitioners have established that the three-edge path is the dominant logistic structure for vegetables traveling to restaurants through wholesaling \cite{38}.\\

We therefore have a network following the characteristic structure defined in Section~\ref{sec:stylizedNets} -- a directed network with four layers $V=\{V_1,V_2,V_3,V_4\}$ and directed edges $(i,j)\in E$  with $i\in V_n$, $j\in V_{n+1},\ n=1,2,3$. The first layer has $V_1 = 452$ nodes representing vegetable production in the German districts and import countries; all other layers have 402 nodes representing only the German districts. The network thus allows for domestic or international production. The nodes on the first three levels are transient nodes $V_Q =\{V_1,V_2,V_3\}$; the nodes on the last level are the absorbing nodes $V_R = V_4$ . The transition matrix $P$ using the German vegetable supply probability matrix $P_V$ as input is then:

\begin{equation*}
P=\left[\begin{array}{cccc}
0 & P_V & 0 & 0\\[2mm]
0 & 0 & P_V & 0\\[2mm]
0 & 0 & 0 & P_R=P_V\\[2mm]
0 & 0 & 0 & I_R\\[2mm]
\end{array}\right].
\end{equation*}

To implement the source estimator on this network, we assume the outbreak began with a producer, i.e. the source is a node $s \in V_1$. Because it was not known what vegetable was causing the infection, we do not assume any prior information regarding the source location and we assume a uniform prior distribution over all feasible sources. We also implement an adjustment to account for the population density of each German district, normalizing the probability of a region reporting an illness by the population density of that region. This ensures the probability of a region reporting an illness is weighted equally.

\subsubsection{Results and discussion}

To evaluate source identification ``in real time,'' we run the source detection methods on data available at the end of each week of the outbreak. Results are reported for our method in combination with the vegetable network model described above and for the Effective Distance method in combination with a network modeling approach reported in \cite{27}, a gravity model network of spatial food transportation based on population statistics. Despite using different network models, our results are comparable because the feasible sources in both network models are the same set of administrative districts in Germany (without including international production). We also note that we evaluated the Effective Distance method in combination with our network, but accuracy was much lower and so here we compare to the results published in \cite{27} (results were only provided for outbreak Weeks 3 -- 9).

We report on accuracy according to two metrics: the rank of the true source, the position of the ordered ranking, and the top-3 distance to the true source, the average distance to the true source in Bienenbu\"uttel from the center of the top three ranked locations.

\begin{table}[b]
\resizebox{\linewidth}{!}{\begin{tabularx}{.55\textwidth}{b|b|b|b|b}
\hline
 \multirow{2}{.2cm}{\\Outbreak Week} & \multicolumn{2}{B|}{Rank of True Source Location} &  \multicolumn{2}{B}{Top-3 Distance from True Source (in km)} \tabularnewline[3mm]
  & This Work & Effective Distance \cite{27}  & This Work & Effective Distance \cite{27}\tabularnewline
  \hline
  1& 38 & -- & 180.0 & --\tabularnewline
  2 & 3 & -- & 148.8 & --\tabularnewline
  3 & 2 & 1 & 83.7 & 71.3\tabularnewline
  4 & 2 & $>$10 & 40.8 & 98.3\tabularnewline
  5 & 1 & 3 & 28.7 & 43.7\tabularnewline
  6 & 1 & 1 & 28.7 & 30.3\tabularnewline
  7& 1 & 1 & 28.7 & 30.3\tabularnewline
  8 & 1 & 5 & 28.7 & 135.0\tabularnewline
  9 & 1 & 2 & 28.7 & 65.0\tabularnewline
  \hline
\end{tabularx}}
\caption{\label{tab:table2}Source identification performance metrics for our method and the Effective Distance method applied to data from the 2011 EHEC O104:H4 outbreak.}
\end{table}

Table~\ref{tab:table2} reports the source identification performance metrics for our method and the Effective Distance method by each week of the outbreak. Our approach is accurate, timely, and consistent, identifying the source district Uelzen in rank 3 by Week 2 of the outbreak, and in the top 2 ranked locations for the remainder of the outbreak. Importantly, though not reported in the table, the district in rank 1 during Weeks 2-4, L\"{u}chow-Dannenberg, is adjacent to Uelzen and its center is as close in distance (km) to the origin farm in Bieneb\"{u}ttel as the center of Uelzen. Furthermore, the consistency of the result in the ordered ranking and the top-3 distance to the true source indicates convergence of the method, signifying a reliable signal for investigators. In comparison, the Effective Distance method is less consistent, identifying the source location within the top 10 ranks in some weeks but not in others, including the critical period around the peak of the outbreak.\\

\begin{figure*}
\begin{center}
\includegraphics[width=0.85\textwidth]{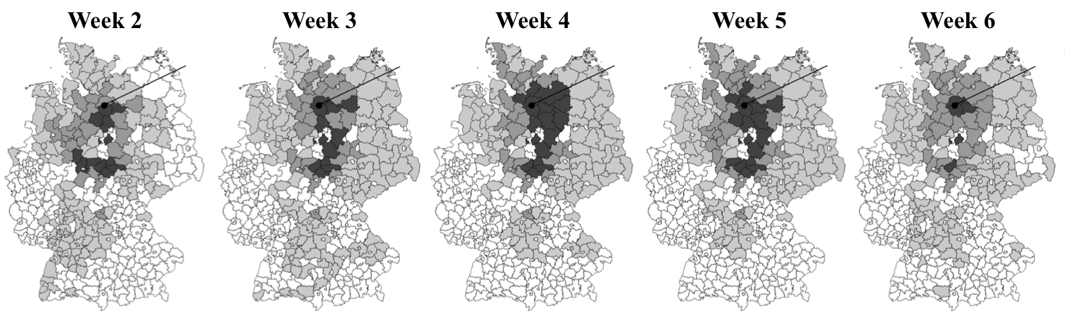}
\end{center}
\caption{\label{fig:EHECresults}Geographical representation of the probability of each of the 402 German administrative districts being the source based on source identification results using our model. Darker shading represents higher probability. The true source in Bienenb\"{u}ttel, district Uelzen, is indicated.}
\end{figure*}

Fig.~\ref{fig:EHECresults} visualizes the probability distribution resulting from applying the source identification method to the case data available at the end of Weeks 2-6, with darker shading representing higher probabilities. The true source in Bieneb\"uttel, district Uelzen, is indicated with the black dot and line. As can be seen in the images, the highest probability locations (also the top ranked locations) frame the outbreak into a small regional area around the true source.

During an outbreak, accurate and timely identification of the source is essential to stem impact on the public. In the context of the EHEC outbreak, this would have meant identifying the source before Week 4, when the outbreak peaked. By developing a source identification approach that accounts for the specific features of foodborne disease transmission and combining it with a network model based on food supply data, we are able to demonstrate significant and timely improvements to source identification on a real case. Our method consistently identifies the source region based on the data available as early as Week 2 of the outbreak, well before the peak. In comparison, the Effective Distance method is inconsistent in identifying the source region, particularly between Weeks 3 and 4, and the standard investigation process (e.g. triangulation) used in the case did not identify the source until Week 6.\\

While our source identification is only as granular as the geographic districts in the network, this information could have been used during the critical early period of the investigation, supplementing conventional methods to help investigators narrow down the list of potential source locations, e.g. to farms located within the black shaded regions in Fig.~\ref{fig:EHECresults}. This also might have prevented investigators from pursuing false leads. This occurred in early stages of the 2011 EHEC outbreak investigation, when investigators wrongly implicated cucumbers produced by a Spanish produce cooperative, wiping out over a month's worth of production and doing lasting damage to the reputation of the Spanish cucumber industry as a whole \footnote{A \$2.54 million settlement was reached in 2015 between the City of Hamburg, whose health officials made the mistaken implication, and the Spanish cooperative \cite{39}}.


\section{\label{sec:Conclusions}Conclusions}

This paper develops a methodology to identify the source of large-scale outbreaks of foodborne disease. We formulate a probabilistic model of the foodborne disease contamination transmission process as an absorbing random walk on a network and derive an estimator for the source location. This is the maximum likelihood source estimator for a diffusion process on a weighted, directed network with absorbing nodes. By formulating the transmission process as a random walk, we are able to develop a novel, computationally tractable solution to the source detection problem that accounts for all possible paths of travel through the network.\\

Existing source location methods for diffusion processes assume that the contamination travels along either the shortest or highest probability paths, which is an unrealistic approximation in the context of many real-world networks, as exemplified by the foodborne disease problem. Thus, the primary contribution of this work is the development of a probabilistically exact source location estimator for large-scale outbreaks that is not limited to tree-like approximations. Application to stylized networks and a real outbreak case demonstrate the benefits of the probabilistically exact estimator in comparison with the state-of-the art approach to source detection of diffusion-type processes on weighted networks. Given the exact food supply network data, our approach shows significant improvements in accuracy and robustness, especially for particular network structures without a unique set of dominant paths. Furthermore, application to real data from the EHEC outbreak demonstrates that our approach is also more reliable, consistently identifying the source location region over the time course of the outbreak. These results suggest our method is closer to adaptation for use by investigators, and could be of help in narrowing down the set of possible sources more quickly.\\

While motivated by the case of foodborne disease, a natural extension of this work is the application to identifying the source of network-based diffusion processes more generally, such as infectious disease spread through global metapopulation-type transport networks or bacterial contaminations spread through water distribution networks. Because the methods developed here consider all possible diffusion trajectories, their application to other real-world problem contexts should demonstrate similar benefits as those shown for foodborne disease, i.e. greater timeliness, accuracy, and reliability in outbreak source localization, especially for non-tree-like network topologies.\\


\begin{acknowledgments}
The authors would like to thank A. Balster for contributions and data sharing relating to the EHEC outbreak case study; E. Polozova for technical assistance; and R. Larson, S. Finkelstein, A. Jacquillat, M. Fuhrmann, and A. Taylor for insightful discussions. This work was developed within the scope of a Robert Wood Johnson Foundation (RWJF) Public Health Services and Systems Research (PHSSR) award and a German Research Foundation (DFG) award. ALH was additionally supported by the Federal Institute for Risk Assessment (BfR) and a Bayer Foundation Award.
\end{acknowledgments}



\begin{thebibliography}{4}%
\makeatletter
\providecommand \@ifxundefined [1]{%
 \@ifx{#1\undefined}
}%
\providecommand \@ifnum [1]{%
 \ifnum #1\expandafter \@firstoftwo
 \else \expandafter \@secondoftwo
 \fi
}%
\providecommand \@ifx [1]{%
 \ifx #1\expandafter \@firstoftwo
 \else \expandafter \@secondoftwo
 \fi
}%
\providecommand \natexlab [1]{#1}%
\providecommand \enquote  [1]{``#1''}%
\providecommand \bibnamefont  [1]{#1}%
\providecommand \bibfnamefont [1]{#1}%
\providecommand \citenamefont [1]{#1}%
\providecommand \href@noop [0]{\@secondoftwo}%
\providecommand \href [0]{\begingroup \@sanitize@url \@href}%
\providecommand \@href[1]{\@@startlink{#1}\@@href}%
\providecommand \@@href[1]{\endgroup#1\@@endlink}%
\providecommand \@sanitize@url [0]{\catcode `\\12\catcode `\$12\catcode
  `\&12\catcode `\#12\catcode `\^12\catcode `\_12\catcode `\%12\relax}%
\providecommand \@@startlink[1]{}%
\providecommand \@@endlink[0]{}%
\providecommand \url  [0]{\begingroup\@sanitize@url \@url }%
\providecommand \@url [1]{\endgroup\@href {#1}{\urlprefix }}%
\providecommand \urlprefix  [0]{URL }%
\providecommand \Eprint [0]{\href }%
\providecommand \doibase [0]{http://dx.doi.org/}%
\providecommand \selectlanguage [0]{\@gobble}%
\providecommand \bibinfo  [0]{\@secondoftwo}%
\providecommand \bibfield  [0]{\@secondoftwo}%
\providecommand \translation [1]{[#1]}%
\providecommand \BibitemOpen [0]{}%
\providecommand \bibitemStop [0]{}%
\providecommand \bibitemNoStop [0]{.\EOS\space}%
\providecommand \EOS [0]{\spacefactor3000\relax}%
\providecommand \BibitemShut  [1]{\csname bibitem#1\endcsname}%
\let\auto@bib@innerbib\@empty
\bibitem [{Note1()}]{Note1}%
  \BibitemOpen
  \bibinfo {note} {In the contagious disease context, weights can be
  interpreted as heterogeneous infection probabilities.}\BibitemShut {Stop}%
\bibitem [{Note2()}]{Note2}%
  \BibitemOpen
  \bibinfo {note} {Prior information, or information external to the network,
  may be available regarding the location of the source. This may come from
  known risk factors, e.g. extreme weather events or sighting of feral wild
  animals, that increase the prevalence of contamination. It may also come from
  expert opinion.}\BibitemShut {Stop}%
\bibitem [{Note3()}]{Note3}%
  \BibitemOpen
  \bibinfo {note} {This is a general formulation that allows paths of different
  lengths to the observations. If no supply network edges exist between nodes
  within a stage or to a previous stage, there will be no cycles in $G$ and $n$
  will be bounded.}\BibitemShut {Stop}%
\bibitem [{Note4()}]{Note4}%
  \BibitemOpen
  \bibinfo {note} {A \protect \$2.54 million settlement was reached in 2015
  between the City of Hamburg, whose health officials made the mistaken
  implication, and the Spanish cooperative \cite {39}}\BibitemShut {NoStop}%
\end{thebibliography}%


\nocite{*}

\begin{thebibliography}{14}
\bibitem[1]{1} Frank, C., Werber, D., Cramer, J. P., Askar, M., Faber, M., an der Heiden, M., et al (2011). ``Epidemic profile of Shiga-toxin-producing Escherichia coli O104: H4 outbreak in Germany.'' \textit{New England Journal of Medicine}, 365(19), 1771-1780.
\bibitem[2]{2} Buchholz, U., Bernard, H., Werber, D., Böhmer, M. M., Remschmidt, C., Wilking, H., et al. (2011). ``German outbreak of Escherichia coli O104: H4 associated with sprouts.'' \textit{New England Journal of Medicine}, 365(19), 1763-1770.
\bibitem[3]{3} Crowe, S. J., Mahon, B. E., Vieira, A. R., Gould, L. H. (2015). ``Vital signs: multistate foodborne outbreaks-United States, 2010-2014.'' \textit{MMWR Morb Mortal Wkly Rep}, 64, 1221-1225.
\bibitem[4]{4} Nsoesie, E. O., Kluberg, S. A., Brownstein, J. S. (2014). ``Online reports of foodborne illness capture foods implicated in official foodborne outbreak reports.'' \textit{Preventive medicine} 67: 264-269.
\bibitem[5]{5} Harris, J. K., et al. (2014). ``Health department use of social media to identify foodborne illness - Chicago, Illinois, 2013-2014.'' \textit{MMWR Morb Mortal Wkly Rep} 63.32: 681-685.
\bibitem[6]{6} Harrison, C., et al. (2014). ``Using online reviews by restaurant patrons to identify unreported cases of foodborne illness - New York City, 2012-2013.'' \textit{MMWR} 63.20: 441-445.
\bibitem[7]{7} Kaufman, J., Lessler, J., Harry, A., Edlund, S., Hu, K., Douglas, J., Thoens, C., Appel, B., Kaesbohrer, A. and Filter, M. (2014). ``A likelihood-based approach to identifying contaminated food products using sales data: performance and challenges.'' \textit{PLoS Comput Biol}, 10(7), p.e1003692.
\bibitem[8]{8} Analytics magazine (2016). ``Grocery store data speeds early detection of food-borne illness."   \url{http://analytics-magazine.org/grocery-store\\-data-speeds-early-detection-of-food-borne-illness/}
\bibitem[9]{9} Food and Drug Administration (FDA) (2001). ``Guide to traceback of fresh fruits and vegetables implicated in epidemiological investigations.'' Rockville (MD): The Division of Emergency and Investigational Operations, Office of Regional Operations, Office of Regulatory Affairs, FDA.
\bibitem[10]{10} Smith, K., Miller, B., Vierk, K., Williams, I., and Hedberg, C. (2015). ``Product Tracing in Epidemiologic Investigations of Outbreaks due to Commercially Distributed Food Items - Utility, Application, and Considerations.'' Council to Improve Foodborne Outbreak Response (CIFOR).
\bibitem[11]{11} Wilkins, M., Julian, E., Kutzko, K., \& Rockhill, S. (2015). ``Outbreak Investigations (Epidemiology).'' Regulatory Foundations for the Food Protection Professional, 105.
\bibitem[12]{12} McEntire, Jennifer, and Tejas Bhatt (2013). ``Pilot Projects for Improving Product Tracing along the Food Supply System - Final Report''. Chicago, IL: Institute of Food Technologists (2013).
\bibitem[13]{13} Painter, J.A., Hoekstra, R.M., Ayers, T., Tauxe, R.V., Braden, C.R., Angulo, F.J., Griffin, P.M. (2013). ``Attribution of foodborne illnesses, hospitalizations, and deaths to food commodities by using outbreak data, United States, 1998- 2008.'' Emerg Infect Dis 2013;19(3):407-15.
\bibitem[14]{14} Shah, D., Zaman, T. (2011). ``Rumors in a network: Who's the culprit?'' \textit{IEEE Transactions on information theory}, 57(8), 5163-5181.
\bibitem[15]{15} Comin, C. H., da Fontoura Costa, L. (2011). ``Identifying the starting point of a spreading process in complex networks.'' \textit{Physical Review E}, 84(5), 056105.
\bibitem[16]{16} Fioriti, V., Chinnici, M. (2012). ``Predicting the sources of an outbreak with a spectral technique.'' \textit{arXiv preprint} arXiv:1211.2333.
\bibitem[17]{17} B. A. Prakash, J. Vreeken, and C. Faloutsos (2014). ``Efficiently spotting the starting points of an epidemic in a large graph,'' \textit{Knowl. Inf. Syst.}, vol. 38, no. 1, pp. 35-59.
\bibitem[18]{18} Pinto, P. C., Thiran, P., Vetterli, M. (2012). ``Locating the source of diffusion in large-scale networks.'' \textit{Physical review letters}, 109(6), 068702.
\bibitem[19]{19} A. Louni and K. P. Subbalakshmi. (2014). ``A two-stage algorithm to estimate the source of information diffusion in social media networks,'' in Proc. \textit{IEEE Conf. Comput. Commun. Workshops (INFOCOM WKSHPS)}, Toronto, ON, Canada, pp. 329-333.
\bibitem[20]{20} Lokhov, A. Y., Me\'{e}zard, M., Ohta, H., \& Zdeborova\'{a}, L. (2014). ``Inferring the origin of an epidemic with a dynamic message-passing algorithm.'' \textit{Physical Review E}, 90(1), 012801.
\bibitem[21]{21} Altarelli, F., Braunstein, A., Dall'Asta, L., Lage-Castellanos, A., Zecchina, R. (2014). ``Bayesian inference of epidemics on networks via belief propagation.'' \textit{Physical review letters}, 112(11), 118701.
\bibitem[22]{22} E. Seo, P. Mohapatra, and T. Abdelzaher, ``Identifying rumors and their sources in social networks,'' in \textit{Proc. SPIE Defense Security Sens.}, Baltimore, MD, USA, 2012.
\bibitem[23]{23} A. Agaskar and Y. M. Lu, ``A fast Monte Carlo algorithm for source localization on graphs,'' in \textit{Proc. SPIE Opt. Eng. Appl.}, San Diego, CA, USA, 2013, Art. no. 88581N.
\bibitem[24]{24} Zhu, K., Ying, L. (2014). ``A robust information source estimator with sparse observations.'' \textit{Computational Social Networks}, 1(1), 3.
\bibitem[25]{25} K. Zhu and L. Ying (2014). ``A robust information source estimator with sparse observations,'' \textit{Comput. Soc. Netw.}, vol. 1, no. 1, p. 1.
\bibitem[26]{26} Brockmann, D., Helbing, D. (2013). ``The hidden geometry of complex, network-driven contagion phenomena.'' \textit{Science}, 342(6164), 1337-1342.
\bibitem[27]{27} Manitz, J., Kneib, T., Schlather, M., Helbing, D., Brockmann, D. (2014). ``Origin detection during food- borne disease outbreaks-a case study of the 2011 EHEC/HUS outbreak in Germany.'' \textit{PLoS currents}, 6.
\bibitem[28]{28} LeBlanc, D. I., Villeneuve, S., Beni, L. H., Otten, A., Fazil, A., McKellar, R., and Delaquis, P. (2015). ``A national produce supply chain database for food safety risk analysis.'' \textit{Journal of Food Engineering}, 147, 24-38.
\bibitem[29]{29} McKellar, Robin C., Denyse I. LeBlanc, Fernando Pérez Rodríguez, and Pascal Delaquis. ``Comparative simulation of Escherichia coli O157: H7 behaviour in packaged fresh-cut lettuce distributed in a typical Canadian supply chain in the summer and winter.'' \textit{Food Control} 35, no. 1 (2014): 192- 199.
\bibitem[30]{30} Pouillot, R., Lubran, M. B., Cates, S. C., \& Dennis, S. (2010). ``Estimating parametric distributions of storage time and temperature of ready-to-eat foods for US households.'' \textit{Journal of food protection}, 73(2), 312-321.
\bibitem[31]{31} Kemeny, J. G., \& Snell, J. L. (1976). \textit{Finite markov chains}. Princeton, NJ: van Nostrand.
\bibitem[32]{32} Newman, M.J. (2003). ``The structure and function of complex networks.'' SIAM Review 45(2):167-256.
\bibitem[33]{33} Weiser, A.A., Gross, S., Schielke, A., Wigger, J.F., Ernert, A., Adolphs, J., Fetsch, A., Müller-Graf, C.,
K\"{a}esbohrer, A., Mosbach-Schulz, O. and Appel, B. (2013). ``Trace-back and trace-forward tools developed ad hoc and used during the EHEC O104: H4 outbreak 2011 in Germany and generic concepts for future outbreak situations.'' \textit{Foodborne pathogens and disease}, 10(3), pp. 263-269.
\bibitem[34]{34} Centers for Disease Control and Prevention (CDC). (2013). ``Outbreak of Escherichia coli O104: H4 infections associated with sprout consumption-Europe and North America, May-July 2011.'' \textit{MMWR. Morbidity and mortality weekly report}, 62(50), 1029.
\bibitem[35]{35} Robert Koch Institute. \href{mailto:SurvStat@RKI.de}{\texttt{SurvStat@RKI.de.}} Accessed December 2016.
\bibitem[36]{36} Balster, A., and Friedrich, H. (2017). ``Dynamic freight flow modelling for risk evaluation in food
supply,'' Transportation Research E, \textit{to appear}.
\bibitem[37]{37} Schubert, M., Kluth, T., Nebauer, G., Ratzenberger, R., Kotzagiorgis, S., Butz, B., Schneider, W., Leible,
M. (2014). Verkehrsverflechtungsprognose 2030 - Los 3: Erstellung der Prognose der deutschlandweiten Verkehrsverflechtungen unter Beru\"{u}cksichtigung des Luftverkehrs. Bundesministerium f\''ur Verkehr und digitale Infrastruktur.
\bibitem[38]{38} Friedrich, H. (2010). Simulation of logistics in food retailing for freight transportation analysis. Doctoral dissertation, Karlsruhe Institute for Technology.
\bibitem[39]{39} The Local (2011). ``Spanish sue Hamburg for E.coli cucumber warning."  \url{https://www.thelocal.de/20111222/39679}
\end{thebibliography}
\end{document}